\documentclass[12pt]{article}
\usepackage{natbib}
\usepackage[left=1.2in,top=1.2in,right=1.2in,bottom=1.2in]{geometry}
\usepackage{amssymb}
\usepackage{amsmath}
\usepackage{MnSymbol}
\usepackage{graphicx}
\usepackage{float}
\usepackage{ragged2e}
\usepackage{setspace}
\usepackage{blindtext}
\usepackage{bbm}
\usepackage{latexsym}
\usepackage{caption}
\usepackage{hyperref}
\usepackage{float}
\usepackage{comment}
\usepackage{dcolumn}
\usepackage{adjustbox}
\usepackage{xcolor}
\usepackage{setspace}
\usepackage{authblk}
\usepackage{subcaption}
\usepackage{booktabs}
\usepackage{multirow}
\usepackage{url}
\usepackage{amsthm}
\usepackage[normalem]{ulem}

\newcommand\independent{\protect\mathpalette{\protect\independenT}{\perp}}
\def\independenT#1#2{\mathrel{\rlap{$#1#2$}\mkern2mu{#1#2}}}

\title{The Local Randomization Framework for Regression Discontinuity Designs: A Review and Some Extensions\footnote{AMS 2010 Subject Classification: 62-02. \textit{Keywords}: regression discontinuity, local randomization, randomization test, assignment mechanism, bandwidth selection}}

\author[1]{Zach Branson\thanks{We would like to thank Luke Keele and Luke Miratrix for insightful comments that led to improvements in this work. Zach Branson would like to thank support from the National Science Foundation Graduate Research Fellowship Program under Grant No. 1144152. Any opinions, findings, and conclusions or recommendations expressed in this material are those of the authors and do not necessarily reflect the views of the National Science Foundation. Fabrizia Mealli's work was supported, in part, by the Dipartimenti Eccellenti 2018-2022 Italian ministerial funds.}}
\author[2]{Fabrizia Mealli}
\affil[1]{Carnegie Mellon University}
\affil[2]{University of Florence}
\date{}

\begin{document}

\maketitle

\noindent
Regression discontinuity designs (RDDs) are a common quasi-experiment in economics and statistics. The most popular methodologies for analyzing RDDs utilize continuity-based assumptions and local polynomial regression, but recent works have developed alternative assumptions based on local randomization. The local randomization framework avoids modeling assumptions by instead placing assumptions on the assignment mechanism near the cutoff. However, most works have focused on completely randomized assignment mechanisms, which posit that propensity scores are equal for all units near the cutoff. In our review of the local randomization framework, we extend the framework to allow for any assignment mechanism, such that propensity scores may differ. We outline randomization tests that can be used to select a window around the cutoff where a particular assignment mechanism is most plausible, as well as methodologies for estimating causal effects after a window and assignment mechanism are chosen. We apply our methodology to a fuzzy RDD assessing the effects of financial aid on college dropout rates in Italy. We find that positing different assignment mechanisms within a single RDD can provide more nuanced sensitivity analyses as well as more precise inferences for causal effects.

\section{Introduction} \label{s:introduction}

Regression discontinuity designs (hereafter RDDs) were first introduced by \cite{thistlethwaite1960regression} and have become a common quasi-experiment in economics, education, medicine, and statistics. RDDs are characterized by an assignment mechanism that is determined by a covariate (or ``running variable'') being above or below a prespecified cutoff. RDDs are similar to observational studies in the sense that treatment assignment is outside researchers' control, but they differ from observational studies in that the treatment assignment mechanism is known. This knowledge of the assignment mechanism can be used to estimate causal effects.

By design, causal effects in RDDs are confounded by the running variable that determines treatment assignment, and thus causal inferences in RDD analyses are typically limited in some way. For example, instead of estimating the average treatment effect (ATE)---the typical estimand in causal inference---researchers usually only attempt to estimate the ATE \textit{at the cutoff} of the running variable, which is identifiable under mild assumptions \citep{hahn2001identification}. By far the most popular methodology for doing this is fitting two local linear regressions---one below the cutoff, one above the cutoff---extrapolating these regressions to the cutoff, and taking the difference between these two extrapolations. This methodology relies on the assumption that the average treatment and control potential outcomes are continuous at the cutoff, and thus this approach falls under what is known as the continuity-based framework \citep{cattaneo2019regression}. \cite{calonico2014robust} introduced a modification to the local linear regression methodology that inflates confidence intervals of treatment effect estimates for more robust causal inferences, and this approach has arguably become the state-of-the-art for estimating the ATE at the cutoff in RDDs. This work has since created a new strand of literature that explores valid causal inferences for treatment effect estimates specifically at the cutoff \citep{imbens2012optimal,calonico2015optimal, calonico2018effect,calonico2019regression,imbens2019optimized}. There are other works that fall under the continuity-based framework but do not use local linear regression---e.g., the use of Bayesian splines \citep{chib2014nonparametric} and Gaussian processes \citep{branson2019nonparametric,rischard2018bayesian}---but these methods are not as popular as local linear regression methods.

Local linear regression methods view the running variable as fixed. Thus, for sharp RDDs, the treatment assignment is fixed as well, because it is a deterministic function of the running variable. In this case, the only stochastic element of an RDD under the continuity-based framework is the repeated sampling of units; thus, causal estimands under this perspective are inherently defined by a hypothetical infinite population. For fuzzy RDDs, the treatment assignment is not fully determined by the running variable, but nonetheless causal inferences under the continuity-based framework are still in reference to a hypothetical infinite population.

An alternative perspective---known as the local randomization framework---views the running variable as random, which introduces a stochasticity to the assignment mechanism in an RDD \citep{li2015evaluating,cattaneo2015randomization,cattaneo2017comparing,cattaneo2019regression}. Under this perspective, it is assumed that there is a window around the cutoff such that treatment assignment for units is as-if randomized; i.e., the assignment mechanism within a window around the cutoff mimics that of a randomized experiment \citep{matteiMealli2016}. This perspective is similar to the perspective taken in the matching literature for observational studies, where it is posited that a matched subset of treatment and control units within an observational study mimics a randomized experiment \citep{rubin2007design,rubin2008objective,rosenbaum2010design}.

Although local regression methods are much more widely used, local randomization methods have several advantages. First, the local randomization framework avoids modeling assumptions, and instead places assumptions on the assignment mechanism for units near the cutoff. Furthermore, inference under the local randomization framework is known to be robust to discrete or continuous running variables, whereas local regression methods may not be applicable in the case of a discrete running variable with a few number of unique values (\citealt{kolesar2018inference}; \citealt[Chapter 3]{cattaneo2018practical}).\footnote{On the other hand, there are some recent continuity-based approaches for discrete running variables; see \cite{li2019regression} and \cite{imbens2019optimized}.} From an applied perspective, local randomization methods also yield an intuitive interpretation: They provide a subset of the data that are meant to mimic a hypothetical randomized experiment, and thus many methods for analyzing randomized experiments can be leveraged. Finally, the local randomization framework is also more conducive to finite-population inference than the continuity-based framework, because the causal estimand is defined by the units at hand, rather than a hypothetical infinite population. Specifically, the causal estimand under the continuity-based framework is the treatment effect at the cutoff for an infinite population, whereas the causal estimand under the local randomization framework is the treatment effect for the units in the sample for which treatment assignment is effectively randomized. This last point is perhaps the most important advantage of the local randomization framework: It allows researchers to easily estimate causal effects beyond the cutoff. This is especially valuable given that the local linear regression literature for RDDs has only recently started exploring how to extrapolate treatment effects beyond the cutoff \citep{mealli2012evaluating,angrist2015wanna,dong2015identifying,cattaneo2018extrapolating}.

Despite these advantages, the local randomization framework has remained largely unnoticed by applied and methodological researchers, and so we believe that it would be valuable to provide a review of the local randomization literature. Indeed, most developments of the local randomization framework have only been published in the past five years, and to the best of our knowledge there is not a review of these methods in the literature. In contrast, local linear regression methods are widely known, and there are several well-known reviews of these methods, including the classic review by \cite{imbens2008regression} and more recent guides such as \cite{cattaneo2017regression} and \cite{cattaneo2019regression}. Although the \cite{cattaneo2019regression} guide does dedicate one section to reviewing the local randomization framework, it would be helpful for the literature to have a comprehensive review of this framework that describes its nuances and discusses possible extensions, and that is what we aim to do here.

In reviewing local randomization methods for RDDs, we will propose new extensions and identify open questions and scope for the literature. Similar to how continuity assumptions are the key ingredient for local linear regression methods, assumptions on the assignment mechanism are the key ingredient for local randomization methods. Despite this, almost all works on local randomization for RDDs have assumed completely randomized assignment mechanisms that are characterized by permutations of the treatment indicator. This is a strong assumption, because it posits that propensity scores are equal for all units near the cutoff. In our review, we will place current local randomization methods within a more general framework, which allows us to easily extend the local randomization framework to any assignment mechanism, such as Bernoulli trials and block randomization, where propensity scores for units near the cutoff are allowed to differ. By relaxing this assumption on the propensity scores, this provides a way to adjust for covariates in an RDD without model specification, which differs from other covariate adjustment methods for RDDs (e.g., \cite{li2015evaluating} and \cite{calonico2019regression}).

Extending the local randomization framework to allow for any assignment mechanism also introduces an interesting trade-off that has not been discussed in the RDD literature. Similar to the local linear regression literature for RDDs, the local randomization literature has put a great deal of attention on how to select the bandwidth, which determines the size of the window in which units are deemed effectively randomized. In our framework, researchers select not only the bandwidth but also the assignment mechanism within the window. Much of the local randomization literature suggests that we would like to achieve the largest bandwidth possible, because it leads to the largest sample size; however, in our framework, a researcher may prefer a smaller bandwidth if they can assume a more precise assignment mechanism. This is analogous to the experimental design liteature, where it may be preferable to conduct a smaller-but-more-precise (e.g., block-randomized) experiment instead of a larger-but-less-precise (e.g., completely randomized) experiment. We will discuss how this trade-off has many important implications for the design and analysis stages of an RDD.

The remainder of the paper is as follows. In Section \ref{s:localRandomization}, we review the local randomization framework for RDDs, and we extend this framework to allow for any assignment mechanism of interest. In Section \ref{s:window}, we review methods for choosing the window around the cutoff for which units are deemed as-if randomized. We also discuss the trade-off between the bandwidth and the assignment mechanism assumed within that bandwidth. After a bandwidth and assignment mechanism are chosen, causal effects can be estimated, and we outline how to do this in Section \ref{s:estimation}. Then, in Section \ref{s:realDataAnalysis}, we demonstrate how to apply the local randomization approach to real data by revisiting a fuzzy RDD---originally analyzed in \cite{li2015evaluating}---that assesses the effects of university grants on dropout rates. In this application, we compare the local randomization approach with local linear regression methods, as well as demonstrate how our framework can give more nuanced analyses than current local randomization approaches by considering multiple assignment mechanisms within an RDD. In Section \ref{s:conclusion}, we conclude.

\section{The Local Randomization Framework for Regression Discontinuity Designs} \label{s:localRandomization}

The core idea behind local randomization methods is that it is hypothesized that there is some window around the cutoff in an RDD such that units are as-if randomized to treatment and control. Then, methods for analyzing randomized experiments can be used to estimate causal estimands within this window.

Unlike local linear regression methods---where stochasticity comes from random sampling from some hypothetical infinite population---under the local randomization framework, stochasticity comes from the assignment mechanism. This idea is more closely aligned with the Rubin Causal Model, which is commonly used for causal inference in experiments and observational studies \citep{holland1986statistics,rubin2005causal}. Furthermore, unlike local linear regression methods, under the local randomization framework, there are many possible causal estimands that can be estimated. In short, the causal estimand under the local randomization framework will depend on the size of the window for which units are as-if randomized: The larger the window, the larger the population for which we can estimate causal effects. Meanwhile, local linear regression methods typically only aim to estimate causal effects \textit{at the cutoff}.

After outlining some notation, we will review the assumptions that local randomization researchers employ to estimate causal effects in RDDs. We will also elaborate on the nature of causal estimands and the assignment mechanism in RDDs, which are the two key concepts for the local randomization framework. The concept of causal estimands is related to how researchers select the ``as-if randomized'' window, which we discuss in Section \ref{s:window}. The concept of assignment mechanisms is related to how researchers estimate causal effects within this as-if randomized window, which we discuss in Section \ref{s:estimation}.

\subsection{Notation} \label{ss:notation}

We follow \cite{imbens2008regression} and discuss RDDs within the potential outcomes framework. Consider $N$ units with potential outcomes $(Y_i(1), Y_i(0))$, where $Y_i(1)$ denotes the outcome of unit $i$ under treatment, and $Y_i(0)$ is analogously defined for control. Let $Z_i$ denote the treatment assignment for unit $i$, where $Z_i = 1$ if unit $i$ is assigned to treatment and 0 otherwise. Let $S_i$ denote the running variable for unit $i$, and let $\mathbf{X}_i$ be a $K$-dimensional vector of other pretreatment covariates. As we discuss further in Section \ref{s:window}, pretreatment covariates besides the running variable are required within the local randomization framework, because these covariates are used to assess if units within a particular window are effectively randomized. Without these additional covariates $\mathbf{X}$, the assumptions discussed in Section \ref{ss:localAssignmentMechanisms} are not testable.

In an RDD, the distribution of the treatment assignment $\mathbf{Z}$ for units with $S \leq s_0$ is different from the distribution of $\mathbf{Z}$ for units with $S > s_0$ for some cutoff $s_0$. The local randomization framework for RDDs focuses analyses on units within a window around the cutoff for which units are effectively randomized to treatment and control. Define $\mathcal{U}_h \equiv \{i: s_0 - h \leq S_i \leq s_0 + h \}$ as the set of units within some window around the cutoff for some \textit{bandwidth} $h$. We choose a symmetric window in part for convenience, and in part because it mirrors previous works on RDDs that fix the bandwidth to be the same above and below the cutoff \citep{imbens2008regression,li2015evaluating}.

\subsection{Causal Estimands: Effects Beyond the Cutoff}

Like most works on causal inference, works on local randomization typically focus on estimating the average causal effect for units in the window $\mathcal{U}_h$, defined as
\begin{align}
	\tau_h \equiv \frac{1}{|\mathcal{U}_h|} \sum_{i \in \mathcal{U}_h} \left[ Y_i(1) - Y_i(0) \right] \label{eqn:tauH}
\end{align}
where $|\mathcal{U}_h|$ denotes the number of units in $\mathcal{U}_h$. Other estimands may also be of interest, such as quantile treatment effects \citep{frandsen2012quantile}, but for simplicity we focus on ATEs.  Importantly, this causal estimand is different from the causal estimand that is estimated by local linear regression methods. Specifically, local linear regression methods aim to estimate the ATE at the cutoff \citep{imbens2008regression}, defined as $\mathbb{E}[Y_i(1) - Y_i(0) | S_i = s_0]$. Because there are no treatment and control units for which $S_i = s_0$, this expectation is taken with respect to some hypothetical infinite population. Meanwhile, the estimand in (\ref{eqn:tauH}) can be written as $\tau_h = \mathbb{E}[Y_i(1) - Y_i(0) | s_0 - h \leq S_i \leq s_0 + h]$, where this expectation is taken with respect to the finite population of units in the sample (we take this finite-population expectation throughout the rest of the paper).

Because finite-population and infinite-population frameworks often yield nearly identical inferential results for randomized experiments \citep{ding2017bridging}, the causal estimand in the local randomization framework is arguably more generalizable than the causal estimand in the continuity-based framework, because inference generalizes to units whose $S_i$ is such that $s_0 - h \leq S_i \leq s_0 + h$ instead of just to units whose $S_i = s_0$. However, this additional generalizability comes at a price: The causal estimand $\tau_h$ is only identifiable given assumptions on the assignment mechanism within the window $\mathcal{U}_h$---namely, units have to actually be as-if randomized within the window. We review these necessary assumptions in the next section, and we discuss how to test these assumptions in Section \ref{s:window}.

\subsection{Local Assignment Mechanism Assumptions for RDDs} \label{ss:localAssignmentMechanisms}

Several assumptions must be placed on the window $\mathcal{U}_h$ in order for $\tau_h$ in (\ref{eqn:tauH}) to be identifiable. In particular, it must be assumed that the assignment mechanism mimicks that of a randomized experiment---namely, that the assignment mechanism is strongly ignorable \citep{rosenbaum1983central} and the Stable Unit Treatment Value Assumption (SUTVA) holds \citep{rubin1980randomization}. In our discussion of these two assumptions, we will closely follow \cite{li2015evaluating}, and we will note how these assumptions are employed elsewhere in the literature. Then, we will turn to particular assumptions that can be placed on the assignment mechanism itself within an RDD.

\subsubsection{SUTVA and Strong Ignorability}

In general, the distribution of the assignment mechanism for an experiment or an observational study may depend on both the potential outcomes and the covariates. Thus, the most general form of the distribution of the assignment mechanism can be written as $P(\mathbf{Z} | \mathbf{Y}(1), \mathbf{Y}(0), \mathbf{X})$. For ease of notation, we focus our discussion on sharp RDDs. In this case, $\mathbf{Z}$ is a deterministic function of $\mathbf{S}$, and thus \citep[Page 1912]{li2015evaluating}:
\begin{align}
	P(\mathbf{Z} = \mathbf{z} | \mathbf{Y}(1), \mathbf{Y}(0), \mathbf{X}) = P(\mathbf{S} \in \boldsymbol{\Lambda} | \mathbf{Y}(1), \mathbf{Y}(0), \mathbf{X}) \label{eqn:rddAssignment}
\end{align}
where $\mathbf{z} \in \{0,1\}^{|\mathcal{U}_h|}$ (i.e., is a particular assignment vector for the units in $\mathcal{U}_h$) and $\boldsymbol{\Lambda} \equiv \{(-\infty, s_0]^{|\mathcal{U}_h|}, (-\infty, s_0]^{|\mathcal{U}_h|-1} \times (s_0, \infty), (s_0, \infty) \times (-\infty, s_0]^{|\mathcal{U}_h|-1}, \dots, (-\infty, s_0] \times (s_0, \infty)^{|\mathcal{U}_h|-1}, (-\infty, s_0]^{|\mathcal{U}_h|-1} \times (s_0, \infty), (s_0, \infty)^{|\mathcal{U}_h|} \}$ (i.e., is the set of possible combinations of each $(S_1,\dots,S_N)$ being above or below the cutoff). Thus, $|\boldsymbol{\Lambda}| = 2^{|\mathcal{U}_h|}$, and $P(\mathbf{S} \in \boldsymbol{\Lambda} | \mathbf{Y}(1), \mathbf{Y}(0), \mathbf{X})$ is the probability distribution on these $2^{|\mathcal{U}_h|}$ combinations. When an RDD is fuzzy instead of sharp, one must consider the distribution of $\mathbf{Z}$ conditional on $\mathbf{S}$, which we elaborate on in Section \ref{s:estimation} when we discuss the analysis stage of an RDD.

Now we will characterize the distribution of the assignment mechanism, (\ref{eqn:rddAssignment}). First, we make the assumption that there is no interference among units and that there are not multiple versions of treatment, commonly known as the Stable Unit Treatment Value Assumption (SUTVA; \citep{rubin1980randomization}). \\ 

\noindent
\noindent\fbox{%
\parbox{\textwidth}{%
\textbf{Local SUTVA}: For each $i \in \mathcal{U}_h$, consider two values $S_i'$ and $S_i''$, where possibly $S_i' \neq S_i''$, corresponding to treatment assignments $Z_i' = I(S_i' > s_0)$ and $Z_i'' = I(S_i'' > s_0)$, where $I(A)$ denotes the indicator function for event $A$. If $Z_i' = Z_i''$, then $Y_i(\mathbf{Z}') = Y_i(\mathbf{Z}'')$.
}
} \\

Local SUTVA is similar to ``Local RD-SUTVA'' in \cite{li2015evaluating}. Local SUTVA states that, for units $i \in \mathcal{U}_h$, the treatment assignment of a unit depends on the running variable only through its being above or below $s_0$, and that the potential outcomes of each unit do not depend on other units' treatment assignment. Importantly, Local SUTVA is also a type of exclusion restriction in that, within the window $\mathcal{U}_h$, it is assumed that there are not different versions of treatment for different values of the running variable. For further discussions about this type of exclusion restriction, see \citet[Section 3]{sekhon2016understanding} and \citet[Section 4]{cattaneo2019regression}.

One additional assumption must be placed on the assignment mechanism in order for $\tau_h$ in (\ref{eqn:tauH}) to be identifiable. Typically, the assignment mechanism of a randomized experiment is assumed to be strongly ignorable, where (1) the potential outcomes are independent of treatment assignment given covariates and (2) there is a non-zero probability of units receiving treatment or control \citep{rosenbaum1983central}. A parallel assumption for an RDD is that the treatment assignment for units around the cutoff is characterized by a strongly ignorable assignment mechanism, which we call Local Strong Ignorability. \\

\noindent
\noindent\fbox{%
\parbox{\textwidth}{%
\textbf{Local Strong Ignorability}: For all $i \in \mathcal{U}_h$,
\begin{align}
(Y_i(1), Y_i(0)) \independent Z_i | \mathbf{X}_i \text{ and } 0 < P(Z_i = 1 | \mathbf{X}_i) < 1 \label{eqn:localStrongIgnorability}
\end{align} 
}
} \\

The assumption that $0 < P(Z_i = 1 | \mathbf{X}_i) < 1$ for all $i \in \mathcal{U}_h$ is similar to the Local Overlap assumption of \cite{li2015evaluating}, and it ensures that there is a non-zero probability on each of the $2^{|\mathcal{U}_h|}$ possible treatment assignments. Furthermore, the independence assumption in Local Strong Ignorability is similar to the Local Randomization assumption of \cite{li2015evaluating}, but it is a weaker assumption in two ways. First, the Local Randomization assumption of \cite{li2015evaluating} posits that the running variable $S_i$ is independent of the potential outcomes, while Local Strong Ignorability posits that only $Z_i$ is independent of the potential outcomes; this is a weaker assumption, because $Z_i$ only depends on $S_i$ through its being above or below the threshold $s_0$. Second, and more importantly, the Local Randomization assumption of \cite{li2015evaluating} posits that this independence holds unconditionally, whereas Local Strong Ignorability only assumes this independence holds conditional on the covariates $\mathbf{X}$. In a similar vein, Local Strong Ignorability is also weaker than unconfoundedness assumptions that have been made in the local randomization literature (e.g., \cite{cattaneo2019regression}) which also assume that this independence between the potential outcomes and the running variable only holds unconditionally. As we discuss in the next section, this relaxation allows us to generalize previous assumptions made in the local randomization literature, such that we can consider more complex assignment mechanisms that may depend on $\mathbf{X}$.

\subsubsection{Types of Local Assignment Mechanisms}

Throughout, we assume that there is a bandwidth $h$ such that Local Strong Ignorability and Local SUTVA hold for all units in the corresponding $\mathcal{U}_h$. Local Strong Ignorability and Local SUTVA are untestable assumptions. However, assuming Local Strong Ignorability and Local SUTVA allows one to test other assumptions about the assignment mechanism for units in $\mathcal{U}_h$. In particular, under Local Strong Ignorability, $P( \mathbf{Z} | \mathbf{Y}(1), \mathbf{Y}(0), \mathbf{X}) = P(\mathbf{Z} | \mathbf{X})$ for units in $\mathcal{U}_h$. In the remainder of this section, we discuss different testable assumptions that can be placed on $P(\mathbf{Z} | \mathbf{X})$ for units in $\mathcal{U}_h$.

To our knowledge, the local randomization literature has focused almost exclusively on one type of assignment mechansim: Complete randomization, where $P(\mathbf{Z} | \mathbf{X})$ corresponds to permutations of $\mathbf{Z}$. In what follows, we discuss several types of assumptions that could be placed on the assignment mechanism within the local randomization framework, but in reality, any strongly ignorable assignment mechanism can be posited within the local randomization framework. Each assumption is increasingly strict, but more strict assumptions---if true---can result in more precise treatment effect estimators. We describe how to test these assumptions in Section \ref{s:window}, and we demonstrate how certain assignment mechanisms lead to more precise treatment effect estimators in Section \ref{s:estimation}.

The least strict assumption that can be placed on the assignment mechanism for units in $\mathcal{U}_h$ is that it follows independent Bernoulli trials. \\

\noindent
\noindent\fbox{%
\parbox{\textwidth}{%
\textbf{Local Bernoulli Trials}: For all $i \in \mathcal{U}_h$,
\begin{align}
P(\mathbf{Z} = \mathbf{z} | \mathbf{X}) = \prod_{i \in \mathcal{U}_h} e(\mathbf{X}_i)^{z_i}[1 - e(\mathbf{X}_i)]^{1 - z_i}, \text{ where } 0 < e(\mathbf{X}_i) < 1 \label{eqn:localBernoulliTrials}
\end{align} 
where $e(\mathbf{X}_i) \equiv P(Z_i = 1 | \mathbf{X}_i)$ is the propensity score for unit $i$.
}
} \\

\cite{li2015evaluating} and \cite{cattaneo2017comparing} considered assumptions similar to Local Bernoulli trials, but they did not allow the trials to depend on other covariates. Thus, the above Local Bernoulli Trials assumption is a generalization of the local randomization assumption in \cite{li2015evaluating} and \cite{cattaneo2017comparing}. Under Local Bernoulli Trials, draws from $P(\mathbf{Z} | \mathbf{X})$ correspond to biased coin flips for the units in $\mathcal{U}_h$ with probabilities corresponding to the propensity scores $e(\mathbf{X})$, which typically need to be estimated (e.g., via logistic regression).

Under the Local Bernoulli Trials assumption, any treatment assignment in $\{0,1\}^{|\mathcal{U}_h|}$ could have plausibly occurred for units in $\mathcal{U}_h$, including the cases where all units are assigned to treatment or all units are assigned to control. However, researchers may not necessarily want to generalize to all of these possible treatment assignments. For example, it is common for analyses of randomized experiments to condition on the number of treated units, the number of units in blocks of covariate strata, and so on. \cite{branson2018randomizationSMMR} discuss how to conduct randomization-based inference for Bernoulli trial experiments conditional on such statistics of interest, and their methods can be used for RDDs if the Local Bernoulli Trials assignment mechanism is assumed.

The remaining locality assumptions we discuss place restrictions on the propensity scores in the Local Bernoulli Trials assumption. The first of these assumptions---Local Complete Randomization---assumes that the propensity scores for all units in $\mathcal{U}_h$ are equal, conditional on the number of units assigned to treatment. \\

\noindent
\noindent\fbox{%
\parbox{\textwidth}{%
\textbf{Local Complete Randomization}: For units $i \in \mathcal{U}_h$:
\begin{equation}
\begin{aligned}
	P(\mathbf{Z} = \mathbf{z} | \mathbf{X}) &= \begin{cases}
	 {|\mathcal{U}_h| \choose N_T}^{-1} \mbox{ if } \sum_{i \in \mathcal{U}_h} z_i = N_T \\
	0 \mbox{ otherwise.}
	\end{cases} \label{eqn:completeRandomizationNT}
\end{aligned}
\end{equation}
}
} \\

Under Local Complete Randomization, draws from $P(\mathbf{Z} | \mathbf{X})$ correspond to random permutations of $\mathbf{Z}^{obs}$ for the units in $\mathcal{U}_h$. This assignment mechanism has been largely the focus of previous local randomization works, including \cite{sales2014limitless}, \cite{cattaneo2015randomization}, \cite{matteiMealli2016}, and \cite{cattaneo2017comparing}.

Now consider the case where the propensity scores of each unit in $\mathcal{U}_h$ are equal within strata of covariates. Say there are $J$ blocks $\mathcal{B}_1,\dots,\mathcal{B}_J$ that divide the covariate space $\mathcal{X}$, such that $\cup_{j=1}^J \mathcal{B}_j = \mathcal{X}$ and $\cap_{j=1}^J \mathcal{B}_j = \emptyset$. Local Block Randomization assumes that the assignment mechanism for units in $\mathcal{U}_h$ mimics that of a block-randomized experiment. \\

\noindent
\noindent\fbox{%
\parbox{\textwidth}{%
\textbf{Local Block Randomization}: For units $i \in \mathcal{U}_h$:
\begin{equation}
\begin{aligned}
	P(\mathbf{Z} = \mathbf{z} | \mathbf{X}) &= P(\mathbf{Z} = \mathbf{z} | \mathcal{B}_1,\dots,\mathcal{B}_J) \\
	&=  \begin{cases}
	 \left[ \prod_{j=1}^J {|\mathcal{U}_h^{(j)}| \choose N_{jT}} \right]^{-1} \mbox{ if } \sum_{i \in \mathcal{U}_h^{(j)}} z_i = N_{jT} \hspace{0.05 in} \forall j=1,\dots,J \\
	0 \mbox{ otherwise.}
	\end{cases} \label{eqn:blockRandomization}
\end{aligned}
\end{equation}
where $\mathcal{U}_h^{(j)} \equiv \{i: i \in \mathcal{U}_h \text{ and } \mathbf{X}_i \in \mathcal{B}_j \}$.
}
} \\

Under Local Block Randomization, draws from $P(\mathbf{Z} | \mathbf{X})$ correspond to random permutations of the observed treatment assignment $\mathbf{Z}^{obs}$ within each block. \cite{cattaneo2015randomization} noted that their assignment mechanism for local RDDs could be generalized to block-randomized designs; Local Block Randomization makes this idea explicit.

In general, researchers can posit any strongly ignorable assignment mechanism for units in $\mathcal{U}_h$. However, for any given $\mathcal{U}_h$, some assignment mechanisms may be more plausible than others. In Section \ref{s:window}, we outline several procedures for testing the plausibility of any strongly ignorable assignment mechanism of interest for an RDD.

\section{How to Select the As-If Randomized Window and Local Assignment Mechanism} \label{s:window}

In order for local randomization methods to yield trustworthy causal inferences, we need to find a bandwidth $h$ such that, within the window $\mathcal{U}_h = \{i: s_0 - h \leq S \leq s_0 + h\}$, it is plausible that Local SUTVA, Local Strong Ignorability, and a particular assignment mechanism $P(\mathbf{Z} | \mathbf{X})$ hold. The first two assumptions are not testable, but researchers can conduct tests to assess if a particular assignment mechanism is plausible within a window $\mathcal{U}_h$. All of these tests require additional pretreatment covariates $\mathbf{X}$. This is in contrast to local linear regression approaches for RDDs, which do not need additional covariates for estimating causal effects. That said, recent research suggests that additional covariates are useful for local linear regression approaches, be it through estimation \citep{calonico2019regression} or via falsification tests \citep{cattaneo2019regression}. The following tests are essentially falsification tests for the local randomization assumptions in Section \ref{s:localRandomization}.

In Section \ref{ss:windowProcedure}, we present procedures for selecting the window $\mathcal{U}_h$ for a particular assignment mechanism. Then, in Section \ref{ss:assignmentMechanismTradeoffs}, we discuss the issue of considering different local assignment mechanisms within a single RDD.

\subsection{Window Selection Procedures} \label{ss:windowProcedure}

All works on local randomization follow the same procedure for selecting the window $\mathcal{U}_h$: \\

\noindent\fbox{%
\parbox{\textwidth}{%
\textbf{Procedure for Selecting the As-If Randomized Window}:
\begin{enumerate}
	\item Posit a strongly ignorable assignment mechanism, denoted by $P(\mathbf{Z} | \mathbf{X})$.
	\item Start by including all units in $\mathcal{U}_h$; i.e., choose $h$ such that $i \in \mathcal{U}_h$ for all $i =1,\dots,N$.
	\item Perform a test to assess if the assignment mechanism from Step 1 is plausible within $\mathcal{U}_h$. 
	\item If the test from Step 3 concludes that the assignment mechanism from Step 1 is plausible within $\mathcal{U}_h$, continue to the analysis stage. Otherwise, decrease the bandwidth $h$, redefine $\mathcal{U}_h = \{i: s_0 - h \leq S_i \leq s_0 + h\}$, and go back to Step 3.
\end{enumerate}
}
} \\

\noindent
Thus, the window is selected to be the largest such that a particular assignment mechanism is plausible, i.e. the units are effectively randomized to treatment and control. It may be the case that there is no window such that the units are effectively randomized. In this case, researchers can posit a different assignment mechanism in Step 1, or they may conclude that local randomization methods cannot be employed.

Although all works on local randomization follow the same above procedure for selecting the window, they differ in the test used in Step 3. Typically, the test in Step 3 will assess if the treatment and control groups within $\mathcal{U}_h$ are balanced with respect to the additional pretreatment covariates $\mathbf{X}$. For example, \cite{li2015evaluating} propose a Bayesian hierarchical modeling approach to test if the covariate mean differences between treatment and control are significantly different from zero. A more common approach is that of \cite{cattaneo2015randomization}, who proposed a randomization-based test for Step 3. After specifying a bandwidth $h$, an assignment mechanism $P(\mathbf{Z} | \mathbf{X})$, and a test statistic $t(\mathbf{Z}, \mathbf{X})$---which measures covariate balance between treatment and control, e.g., the covariate mean difference---this test proceeds as follows: \\

\noindent\fbox{%
\parbox{\textwidth}{%
\textbf{Generalization of $\alpha$-level Randomization Test of \cite{cattaneo2015randomization}}:
\begin{enumerate}
	\item Generate random draws $\mathbf{z}^{(1)},\dots,\mathbf{z}^{(M)} \sim P(\mathbf{Z} | \mathbf{X})$ for units $i \in \mathcal{U}_h$. $M$ should be large enough to well-approximate the randomization distribution of the test statistic (e.g., $M = 1,000$).
	\item For each covariate $k=1,\dots,K$, compute the randomization-based $p$-value:
	\begin{align}
		p_k = \frac{\sum_{m=1}^M \mathbb{I}(t(\mathbf{z}^{(m)}, \mathbf{X}^{(k)}) \geq t_k^{obs})}{M + 1} \label{eqn:randPValue}
	\end{align}
	where $t^{obs} \equiv t \left(\mathbf{z}^{obs}, \mathbf{X}^{(k)} \right)$ and $\mathbf{X}^{(k)}$ is the vector of values for the $k$th covariate.
	\item Define $p_{min} = \min \{p_k\}_{k=1,\dots,K} $
	\item Reject the plausibility of $P(\mathbf{Z} | \mathbf{X})$ if $p_{min} < \alpha$.
\end{enumerate}
	}
	} \\

The test of \cite{cattaneo2015randomization} only considered Local Complete Randomization as an assignment mechanism, and thus Step 1 consisted of generating permutations of $\mathbf{Z}^{obs}$. In the above, we have generalized this test to consider any assignment mechanism. For example, if Local Block Randomization is posited, then Step 1 consists of generating permutations of $\mathbf{Z}^{obs}$ within blocks.

The above procedure assesses if covariates are balanced to the degree we'd expect them to be balanced from a randomized experiment, where balance is measured by the test statistic $t(\mathbf{Z}, \mathbf{X})$. Researchers can specify a different test statistic for each covariate---for example, \cite{cattaneo2015randomization} used the covariate mean differences---and then conservatively take the minimum $p$-value across these test statistics. Alternatively, researchers can use an omnibus measure of covariate balance to produce one $p$-value, as we will do in Section \ref{s:realDataAnalysis}. Importantly, $t(\mathbf{Z}, \mathbf{X})$ is not a function of the outcomes, in order to prevent researchers from biasing results when determining the assignment mechanism \citep{rubin2007design,rubin2008objective}. Different test statistics will result in different levels of statistical power, as discussed in \cite{cattaneo2015randomization} for randomization tests within RDDs and \cite{rosenbaum2002observational} and \cite{imbens2015causal} for randomization tests in general.

As discussed in \cite{cattaneo2015randomization}, units in $\mathcal{U}_h$ are deemed effectively randomized if the above procedure fails to reject the null hypothesis of as-if randomization. Thus, unlike most hypothesis testing settings, in this context we are more concerned about making a Type II error (i.e., failing to the reject the null hypothesis when it is actually false) than about making a Type I error. For this reason, \cite{cattaneo2015randomization} recommend selecting a slightly larger $\alpha$ than $\alpha = 0.05$ (based on power calculations, they recommend $\alpha = 0.15$), and to be conservative, the rejection rule is based on the minimum $p$-value, as shown in Step 3 above. Furthermore, when the above test is used for several bandwidths $h$ (which is likely to be the case, given the ``Procedure for Selecting the As-If Randomized Window'' above), \cite{cattaneo2015randomization} recommend \textit{not} making multiple hypothesis testing corrections, again for Type II error considerations. Similar testing procedures have been used for local randomization in RDDs \citep{sales2014limitless,matteiMealli2016,cattaneo2017comparing,cattaneo2019regression} and matching for observational studies \citep{hansen2008essential,hansen2008covariate,gagnon2016classification,branson2018my}. Results from this testing procedure can also be summarized graphically, as we demonstrate in our application in Section \ref{s:realDataAnalysis}.

\subsection{The Trade-Off between Bandwidths and Assignment Mechanisms} \label{ss:assignmentMechanismTradeoffs}

The previous subsection outlines procedures for selecting the window $\mathcal{U}_h$ for a particular local assignment mechanism. To our knowledge, all works in the literature have only considered a single assignment mechanism---usually local complete randomization, defined in (\ref{eqn:completeRandomizationNT})---when using these procedures. However, under our framework for local randomization, any local assignment mechanism may be posited. At the expense of slightly abusing notation, consider $A$-many local assignment mechanisms $P_1(\mathbf{Z} | \mathbf{X}), \dots, P_A(\mathbf{Z} | \mathbf{X})$ that a researcher posits for an RDD, and say that they have found corresponding bandwidths $h_1,\dots,h_A$ for which these assignment mechanisms are deemed ``plausible'' according to procedures discussed in Section \ref{ss:windowProcedure}. Each bandwidth and assignment mechanism pair will suggest a particular analysis for estimating causal effects, as we outline in Section \ref{s:estimation}. How should we come to terms with $A$-many different analyses for a single RDD, all of which are deemed ``plausible'' by procedures used in the local randomization literature? This presents an interesting quandary and trade-off between bandwidths and assignment mechanisms that, to our knowledge, has not been previously discussed in the literature.

First, it is important to note that, given bandwidths $h_1,\dots,h_A$, each analysis is estimating a different causal estimand. This is fundamentally different from the continuity-based framework, where different bandwidths still correspond to the same causal estimand of the ATE at the cutoff. Throughout, we have been assuming that the causal estimand for a bandwidth $h$ is the local ATE $\tau_h$, defined in (\ref{eqn:tauH}). Thus, the causal estimand for each analysis is $\tau_{h_1},\dots,\tau_{h_A}$, which may well be different quantities. Of course, these quantities are nonetheless very related to each other: For example, if $h_1 < h_2$, then it is important to note that the window $\mathcal{U}_{h_1}$ is a subset of $\mathcal{U}_{h_2}$. Thus, the ATE $\tau_{h_2}$ can be written as
\begin{equation}
\begin{aligned}
	\tau_{h_2} &= \frac{\sum_{i \in \mathcal{U}_{h_2} } [Y_i(1) - Y_i(0)]}{|\mathcal{U}_{h_2}|} \\
	&= \frac{\sum_{i \in \mathcal{U}_{h_1} } [Y_i(1) - Y_i(0)] + \sum_{i \in \{ \mathcal{U}_{h_2} \setminus \mathcal{U}_{h_1} \} } [Y_i(1) - Y_i(0)] }{|\mathcal{U}_{h_2}|} \\
	&= \frac{|\mathcal{U}_{h_1}| \cdot \tau_{h_1} + |\mathcal{U}_{h_2} \setminus \mathcal{U}_{h_1}| \cdot \tau_{ \{h_2 \setminus h_1 \}}}{|\mathcal{U}_{h_2}|}
\end{aligned}
\end{equation}
where the notation $A \setminus B$ denotes the set $A$ minus the set $B$, and $\tau_{ \{h_2 \setminus h_1 \}}$ denotes the ATE for units in $\mathcal{U}_{h_2} \setminus \mathcal{U}_{h_1}$. In other words, $\tau_{h_2}$ is a weighted average of the ATE for units in $\mathcal{U}_{h_1}$ and the ATE for units in $\mathcal{U}_{h_2}$ but not $\mathcal{U}_{h_1}$. Thus, if $\tau_{h_2}$ and $\tau_{h_1}$ are estimated to be very different from each other, this suggests something peculiar about the ATE for units in $\mathcal{U}_{h_2} \setminus \mathcal{U}_{h_1}$ (which are also the units farthest from the cutoff). Such differences in ATEs could be the source of treatment effect heterogeneity. For this reason, the local randomization literature has recommended conducting sensitivity analyses, where the treatment effect is estimated for several ``plausible'' bandwidths to assess if causal effect estimates are robust across different bandwidth choices \citep{li2015evaluating,cattaneo2015randomization,cattaneo2017comparing}.

Under our framework, we suggest a similar sensitivity analysis, but we recommend reporting several causal effect estimates for different bandwidth and assignment mechanism pairs. Thus, sensitivity analyses currently seen in the literature can be viewed as a special case of our proposed sensitivity analysis, where the assignment mechanism is the same across all plausible bandwidth and assignment mechanism pairs. However, we believe that it is important to consider different assignment mechanisms, because more precise assignment mechanisms can lead to more precise inference for causal effects. For example, block randomization is considered to be a more precise assignment mechanism than complete randomization, in the sense that it tends to provide narrower confidence intervals for causal effects. At the same time, more precise assignment mechanisms may only be deemed plausible for smaller bandwidths. In short, it may be the case that some causal estimands can be more precisely estimated than others, and that would be enlightening information to have in a sensitivity analysis.

To understand how such a sensitivity analysis would work, first we need to review how causal effects are estimated under the local randomization framework after the window and local assignment mechanism are chosen. We do this in Section \ref{s:estimation}. Then, in Section \ref{s:realDataAnalysis}, we demonstrate local randomization approaches on a university grant dataset originally analyzed in \cite{li2015evaluating}. In that section, we compare and contrast continuity-based and local randomization approaches, and we also show how considering different assignment mechanisms under our framework can lead to more nuanced results and insightful sensitivity analyses than what has previously been discussed in the local randomization literature.

\section{How to Estimate Causal Effects after Selecting the Window and Local Assignment Mechanism} \label{s:estimation}

Section \ref{s:localRandomization} outlined locality assumptions resesarchers can posit for an RDD. These assumptions assert that---for some bandwidth $h$---the treatment assignment for units in a window $\mathcal{U}_h$ follows an assignment mechanism $P(\mathbf{Z} | \mathbf{X})$. Meanwhile, Section \ref{s:window} outlined procedures for selecting a window $\mathcal{U}_h$ for which a particular assignment mechanism is plausible. In this section, we discuss how causal effects can be estimated under the local randomization framework after a window and assignment mechanism have been specified.

After units in $\mathcal{U}_h$ have been deemed effectively randomized according to an assignment mechanism $P(\mathbf{Z} | \mathbf{X})$, the analysis of the RDD should mimic the analysis of a randomized experiment where treatment assignment follows the distribution $P(\mathbf{Z} | \mathbf{X})$ specified by the researcher. There is a huge literature on randomization-based, Neymanian, and Bayesian modes of inference for analyzing such randomized experiments; see \cite{imbens2015causal} for a general review of how to analyze randomized experiments for different assignment mechanisms. Thus, one benefit of the local randomization aproach is that this literature can be readily leveraged for estimating causal effects in RDDs. For example, \cite{cattaneo2015randomization} outlined a randomization-based approach based on \cite{rosenbaum2002observational,rosenbaum2010design} that is particularly useful for small samples, and \cite{li2015evaluating} outlined a Bayesian approach utilizing principal stratification for analyzing a fuzzy RDD (which we will revisit in Section \ref{s:realDataAnalysis}).

To complement the literature, in this section we outline how to conduct Neymanian inference for RDDs under the local randomization framework. To our knowledge, this mode of inference has not been outlined for RDDs under the local randomization framework, but it is straightforward given the large literature on Neymanian inference for randomized experiments.\footnote{Standard methods for analyzing RDDs---e.g., local linear regression---may also be considered Neymanian modes of inference for RDDs. However, such methods are from a continuity-based perspective, and here we outline a Neymanian mode of inference from a local randomization perspective.} To mirror our recommendation that researchers consider multiple assignment mechanisms within a single RDD, we will outline how to analyze an RDD under Local Complete Randomization (defined in (\ref{eqn:completeRandomizationNT})) and Local Block Randomization (defined in (\ref{eqn:blockRandomization})). We will show how to do this for fuzzy RDDs, where sharp RDDs can be considered a special case. Because there is an equivalence between fuzzy RDDs and sharp RDDs with noncompliance \citep{hahn2001identification,imbens2008regression,li2015evaluating}, we will use theory from analyzing randomized experiments with noncompliance to establish our results. Thus, the remainder of this section can also act as a guide for how researchers can leverage theory from randomized experiments to establish new results for analyzing RDDs. For example, recent research has established Neymanian-style results for analyzing experiments with complex assignment mechanisms like cluster-randomization \citep{middleton2015unbiased}, rerandomization \citep{li2018asymptotic}, and ``hybrid designs'' that mix matched pairs and blocks \citep{pashley2017insights}. We believe a promising area of future research is establishing how these complex assignment mechanisms can be utilized in RDDs.

\subsection{How to Analyze a Fuzzy RDD assuming Local Complete Randomization} \label{ss:howToCR}

First we will outline how to analyze any particular $\mathcal{U}_h$ assuming Local Complete Randomization, and then we will extend this analysis to Local Block Randomization. Specifically, we will establish point estimates and conservative Neymanian confidence intervals for the local ATE $\tau_h$ for compliers, which we will define shortly.

Under the local randomization framework, the analysis for a fuzzy RDD under Local Complete Randomization is identical to the analysis of a completely randomized experiment with noncompliance. For ease of exposition, we will assume one-sided noncompliance, where every unit in the control group actually receives control but some units in the treatment group may not actually receive treatment. Such a setup matches the application we will analyze in Section \ref{s:realDataAnalysis}. We will follow the methodology outlined by \citet[Chaper 23]{imbens2015causal} for analyzing randomized experiments with one-sided noncompliance, but adapted to the context of our local randomization framework for RDDs. Our results can be extended to the more general case of two-sided noncompliance by following \citet[Chaper 24]{imbens2015causal}.

For this section, let $Z_i = \mathbb{I}(S_i > s_0)$ denote the treatment assignment or ``encouragement'' to treatment in the RDD. Furthermore, let $W_i(Z_i)$ denote the actual binary treatment receipt in the RDD when assigned to treatment $Z_i$, and let $W_i = Z_i W_i(1) + (1-Z_i) W_i(0)$ denote the observed treatment receipt. As is common practice, we are particularly interested in estimating the local ATE for compliers, which is the set of units $\mathcal{C}$ such that $W_i(Z_i) = Z_i$. Then, the local ATE for compliers is defined as
\begin{align}
	\tau_h^c = \sum_{i \in \mathcal{U}_h \cap \mathcal{C}} [Y_i(1) - Y_i(0)]
\end{align}
Although we typically cannot identify the set of compliers $\mathcal{C}$, we can identify $\tau_h^c$ under the additional assumption of the exclusion restriction, which states that the treatment assignment $Z$ only affects the outcomes through the treatment receipt $W$ \citep{angrist1996identification}. For example, the exclusion restriction allows us to state that for never-takers (i.e., units such that $W_i(1) = W_i(0) = 0$), $Y_i(1) = Y_i(0)$, because never-takers do not receive treatment when $Z_i = 1$ or $Z_i = 0$.\footnote{This is because the potential outcomes $Y_i(1)$ and $Y_i(0)$ are indexed by $Z_i$, not $W_i$.} Furthermore, under one-sided noncompliance, $W_i(0) = 0$ for all units, which is the setting we will consider in Section \ref{s:realDataAnalysis}.

Under the exclusion restriction and one-sided noncompliance, the point estimate of $\tau_h^c$ is the ratio of the intention-to-treat effect on the outcome and the intention-to-treat effect on the treatment receipt indicator:
\begin{align}
	\widehat{\text{ITT}}_Y^{(h)} &= \frac{\sum_{i: Z_i = 1, i \in \mathcal{U}_h} Y_i}{N_T^{(h)}}  - \frac{\sum_{i: Z_i = 0, i \in \mathcal{U}_h} Y_i}{N_C^{(h)}} \label{eqn:ittY} \\
	\widehat{\text{ITT}}_W^{(h)} &= \frac{\sum_{i: Z_i = 1, i \in \mathcal{U}_h} W_i}{N_T^{(h)}}  - \frac{\sum_{i: Z_i = 0, i \in \mathcal{U}_h} W_i}{N_C^{(h)}} \label{eqn:ittW}
\end{align}
where $N_T^{(h)} = \sum_{i \in \mathcal{U}_h} Z_i$ and $N_C^{(h)} = \sum_{i \in \mathcal{U}_h} (1 - Z_i)$. Then, the point estimate of $\tau_h^c$ is
\begin{align}
	\hat{\tau}_h^c = \frac{\widehat{\text{ITT}}_Y^{(h)}}{\widehat{\text{ITT}}_W^{(h)}} \label{eqn:tauHEstimate}
\end{align}
To quantify uncertainty in this point estimate, using results from \citet[Page 531]{imbens2015causal}, we have that
\begin{align}
	\text{Var}\left(\hat{\tau}_h^c\right) &= \frac{\text{Var}(\widehat{\text{ITT}}^{(h)}_Y)}{(\widehat{\text{ITT}}_W^{(h)})^2} + \frac{(\widehat{\text{ITT}}_Y^{(h)})^2}{(\widehat{\text{ITT}}_W^{(h)})^4} \text{Var}(\widehat{\text{ITT}}^{(h)}_W) - 2 \frac{\widehat{\text{ITT}}_Y^{(h)}}{(\widehat{\text{ITT}}_W^{(h)})^3} \text{Cov}(\widehat{\text{ITT}}_Y^{(h)}, \widehat{\text{ITT}}_W^{(h)}) \label{eqn:varianceTauH}
\end{align}
Thus, a Neymanian $\alpha$-level confidence interval for $\tau_h^c$ assuming Local Complete Randomization is
\begin{align}
	\hat{\tau}_h^c \pm  z_{\alpha/2} \sqrt{\widehat{\text{Var}}\left(\hat{\tau}_h^c \right)}
\end{align}
where $z_a$ denotes the $a$th quantile of the standard Normal distribution, and $\widehat{\text{Var}}(\hat{\tau}_h)$ is obtained by plugging in estimates for $\text{ITT}_Y^{(h)}$, $\text{ITT}_W^{(h)}$, $\text{Var}(\text{ITT}_Y^{(h)})$, $\text{Var}(\text{ITT}_W^{(h)})$, and $\text{Cov}(\text{ITT}_Y^{(h)}, \text{ITT}_W^{(h)})$ into (\ref{eqn:varianceTauH}). The point estimates for $\text{ITT}_Y^{(h)}$ and $\text{ITT}_W^{(h)}$ are shown in (\ref{eqn:ittY}) and (\ref{eqn:ittW}), respectively. Using results from \citet[Chapter 23]{imbens2015causal}, conservative estimates for $\text{Var}(\widehat{\text{ITT}}_Y^{(h)})$ and $\text{Var}(\widehat{\text{ITT}}_W^{(h)})$ are
\begin{align}
	\widehat{\text{Var}}(\widehat{\text{ITT}}_Y^{(h)}) &= \frac{\sum_{i: Z_i = 1, i \in \mathcal{U}_h} (Y_i - \bar{Y})^2}{N_T^{(h)}(N_T^{(h)} - 1)} + \frac{\sum_{i: Z_i = 0, i \in \mathcal{U}_h} (Y_i - \bar{Y})^2}{N_C^{(h)}(N_C^{(h)} - 1)} \\
	\widehat{\text{Var}}(\widehat{\text{ITT}}_W^{(h)}) &= \frac{\sum_{i: Z_i = 1, i \in \mathcal{U}_h} (W_i - \bar{W})^2}{N_T^{(h)}(N_T^{(h)} - 1)} + \frac{\sum_{i: Z_i = 0, i \in \mathcal{U}_h} (W_i - \bar{W})^2}{N_C^{(h)}(N_C^{(h)} - 1)}
\end{align}
i.e., the standard Neymanian variance estimators for the average intention-to-treat effect on the outcome and treatment receipt. Finally, using results from \citet[Page 541]{imbens2015causal}, an estimate for $\text{Cov}(\widehat{\text{ITT}}_Y^{(h)}, \widehat{\text{ITT}}_W^{(h)})$ is
\begin{align}
	\widehat{\text{Cov}}(\widehat{\text{ITT}}_Y^{(h)}, \widehat{\text{ITT}}_W^{(h)}) = \frac{1}{N_T^{(h)} (N_T^{(h)} - 1)} \sum_{i: Z_i = 1, i \in \mathcal{U}_h} (Y_i - \bar{Y})( W_i - \bar{W} ) 
\end{align}
i.e., the sample covariance between $Y_i$ and $W_i$ in the treatment group.

\subsection{How to Analyze a Fuzzy RDD assuming Local Block Randomization} \label{ss:howToBlock}

Now let $\mathcal{B}^{(h)}_1,\dots,\mathcal{B}^{(h)}_J$ denote the blocks of units within $\mathcal{U}_h$ for a particular instance of the Local Block Randomization assumption. Using standard results about block randomized experiments (e.g., \cite{miratrix2013adjusting}), we can leverage the results from the previous subsection to construct a Neymanian point estimate and confidence interval assuming Local Block Randomization.

Let $\hat{\tau}_h^{c(j)}$ denote the point estimate for $\tau_h^c$ within block $\mathcal{B}_j^{(h)}$. In other words, $\hat{\tau}_h^{c(j)}$ is computed as $\hat{\tau}_h^c$ defined in (\ref{eqn:tauHEstimate}), but only using units in $\mathcal{B}_j^{(h)}$. Then, an estimate for $\tau_h^c$ under Local Block Randomization can be constructed by taking a weighted average of the $\hat{\tau}_h^{c(j)}$, where the weights correspond to the sizes of the blocks:
\begin{align}
	\hat{\tau}_h^{c(\text{block})} &= \frac{\sum_{j=1}^J |\mathcal{B}_j^{(h)}| \cdot \hat{\tau}_h^{c(j)}}{\sum_{j=1}^J |\mathcal{B}_j^{(h)}|} 
\end{align}
Similarly, an estimate for the variance of $\hat{\tau}_h^{c(\text{block})}$ is
\begin{align}
	\widehat{\text{Var}} \left(\hat{\tau}_h^{c(\text{block})} \right) &= \frac{\sum_{j=1}^J |\mathcal{B}_j^{(h)}|^2 \cdot \widehat{\text{Var}}(\hat{\tau}_h^{c(j)})}{\left( \sum_{j=1}^J |\mathcal{B}_j^{(h)}| \right)^2} 
\end{align}
where $\widehat{\text{Var}}(\hat{\tau}_h^{c(j)})$ is computed as $\widehat{\text{Var}}(\hat{\tau}_h^c)$ defined in (\ref{eqn:varianceTauH}), but only using units in $\mathcal{B}_j^{(h)}$. Thus, a Neymanian $\alpha$-level confidence interval for $\tau_h$ assuming Local Block Randomization is
\begin{align}
	\hat{\tau}_h^{c(\text{block})} \pm  z_{\alpha/2} \sqrt{\widehat{\text{Var}}\left(\hat{\tau}_h^{c(\text{block})}\right)}
\end{align}

In what follows, we will utilize these point estimates and confidence intervals to analyze a real fuzzy RDD.

\section{Real Data Analysis: Revisiting the Effects of University Grants on Student Dropout} \label{s:realDataAnalysis}

\cite{li2015evaluating} examined a dataset of first-year students at the University of Pisa and University of Florence from 2004 to 2006. In Italy, state universities offer financial grants to students whose families are deemed low-income. Determining whether a student is from a low-income household involves an economic measure that combines information from tax returns, property, and family size. This economic measure is the running variable $S_i$. Students were deemed eligible for state university grants if $S_i \leq 15,000$ euros. \cite{li2015evaluating} define $Z_i$ as the ``eligibility'' or ``encouragement'' for treatment; i.e., $Z_i \equiv \mathbb{I}(S_i \leq 15,000)$.

In addition to the running variable $S_i$ and eligibility $Z_i$, the data include dropout status at the end of the academic year (the binary outcome, $Y_i$), whether or not students applied for a grant ($A_i$), whether or not students received a grant ($W_i$) and the following covariates ($\mathbf{X}_i$): gender, high school grade, high school type (four categories), major in university (six categories), year of enrollment (2004, 2005, or 2006), and university (Pisa or Florence). The goal of the analysis was to assess if offering financial aid through grants lessened students' chances of dropping out of unviersity after their first year. Exploratory data analyses and summary statistics about this student population can be found in \cite{li2015evaluating}.

There were some students who were eligible for a grant (i.e., $Z_i = 1$) that nonetheless did not apply for a grant (i.e., $A_i = 0$ and thus $W_i = 0$). Furthermore, there were students who were ineligible for a grant that nonetheless applied for one---this is likely because the economic measure $S_i$ is complicated for students to measure on their own, and thus they do not know their exact value of $S_i$ when they apply. In short, students can only receive a grant if they are eligible and apply for one: $\{A_i = 1 \text{ and } Z_i = 1\} \leftrightarrow W_i = 1$.

When one views $W_i$ (grant receipt status) as the treatment assignment, this setup is a fuzzy RDD, because $P(W_i = 1 | Z_i = 0) = 0$ but $0 < P(W_i = 1 | Z_i = 1) < 1$. Equivalently, one can view $Z_i$ (eligibility to receive the grant) as the treatment assignment or ``encouragement'' but $W_i$ as the actual receipt of treatment, and thus this setup is a sharp RDD with issues of noncompliance.

\cite{li2015evaluating} used a Bayesian procedure to first find a balanced subpopulation $\mathcal{U}_h$, and then used an additional Bayesian principal stratification approach to estimate the treatment effect within $\mathcal{U}_h$ while accounting for the application status, $A_i$. As discussed in \cite{li2015evaluating}, the application status can provide additional information about the nature of causal effects. However, we will ignore application status in our analysis in order to mimic more standard fuzzy RDD analyses, because the main purpose of this section is to outline how to conduct a local randomization analysis for a fuzzy RDD.

For ease of discussion, we will refer to $Z_i$ (eligibility) as the treatment assignment and recognize that there is one-sided noncompliance with this treatment assignment. The noncompliance is one-sided because $Z_i = 0 \rightarrow W_i = 0$ but it may be the case that $Z_i = 1$ while $W_i = 0$. First, we will demonstrate how to estimate causal effects under continuity-based assumptions using state-of-the-art local linear regression approaches. Then, we will demonstrate how to estimate causal effects under local randomization assumptions.

\subsection{Analyzing the Data using Local Linear Regression} \label{ss:llrAnalysis}

\begin{figure}[H]
    \centering
    \begin{subfigure}[t]{0.45\textwidth}
        \centering
        \includegraphics[scale=0.4]{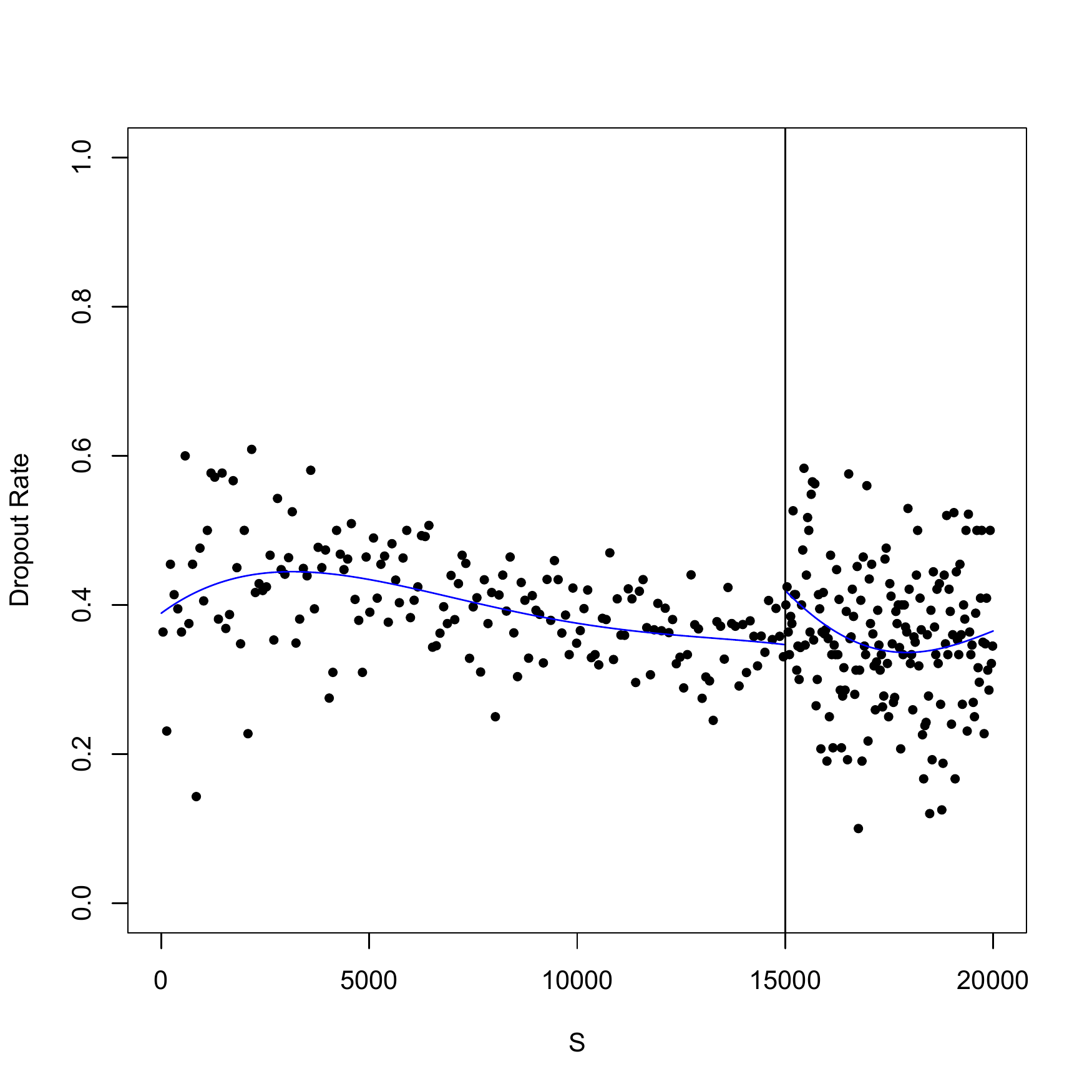}
        \caption{Mean function for $Y$.}
        \label{fig:localLinearRegressionYPlot}
    \end{subfigure}%
    ~
    \begin{subfigure}[t]{0.45\textwidth}
        \centering
        \includegraphics[scale=0.4]{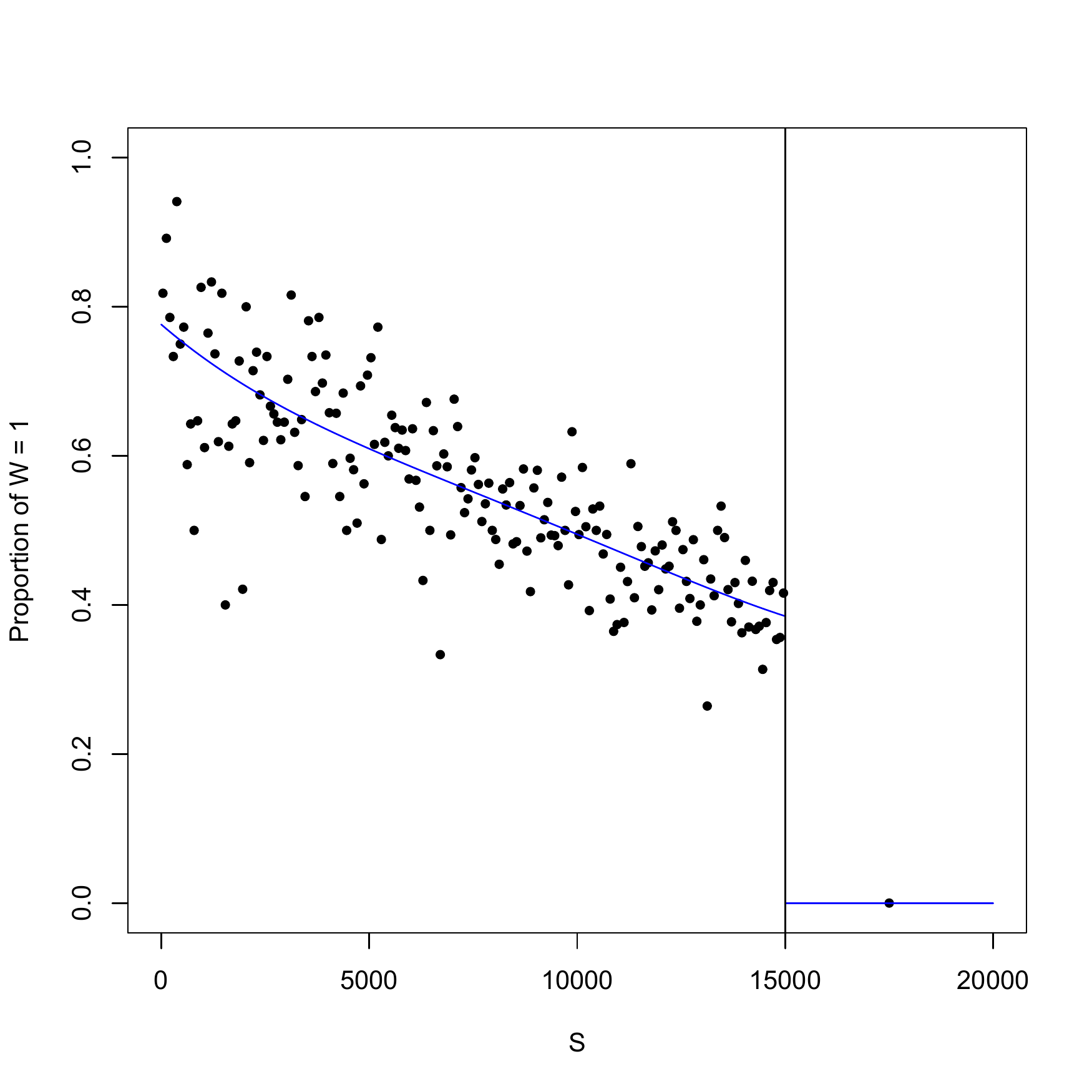}
        \caption{Mean function for $W$.}
        \label{fig:localLinearRegressionWPlot}
    \end{subfigure}%

    \caption{Plots of the estimated mean functions for the outcome $Y$ and treatment receipt $W$ using local linear regression. }
    \label{fig:localLinearRegressionPlots}
\end{figure}

The local linear regression approach estimates the complier ATE \textit{for units at the cutoff} in a way analogous to the estimator (\ref{eqn:tauHEstimate}), where the quantities $\widehat{\text{ITT}}_Y^{(h)}$ and $\widehat{\text{ITT}}_W^{(h)}$ are estimated by fitting two local linear regressions---one below the cutoff and one above the cutoff---for the outcome and treatment receipt indicator, respectively. To provide some visual intuition for this approach, Figure \ref{fig:localLinearRegressionPlots} shows an aggregated, binned version of the data, as well as the mean functions that local linear regression estimates on either side of the cutoff for the outcome $Y$ and the treatment receipt $W$. We implemented these plots in the \texttt{rdrobust} \texttt{R} package \citep{calonico2015optimal,calonico2015rdrobust}.

In Figure \ref{fig:localLinearRegressionYPlot}, we can see that students who marginally do not receive the grant (right-hand side of the cutoff) are estimated to have a higher dropout rate than students who marginally do receive the grant. In Figure \ref{fig:localLinearRegressionWPlot}, we can see that no student with $S > 15000$ receives the grant, but only a fraction of eligible students with $S \leq 15000$ receive the grant (because only a fraction of them apply). Meanwhile, Table \ref{tab:llrEstimates} shows the point estimate, $p$-value, and confidence interval for the ATE at the cutoff from local linear regression methods. The first row of Table \ref{tab:llrEstimates} uses the popular approach of \cite{calonico2014robust}---implemented in the \texttt{rdroubst} \texttt{R} package\footnote{Implementing local linear regression requires determining a bandwidth for the local regression. We used an MSE-optimal bandwidth, which is the default choice in the \texttt{rdrobust} package. Simulation studies (e.g., \cite{branson2019nonparametric}) have found that results for large datasets like the grant dataset tend to be similar across different bandwidth choices, and so we do not consider other bandwidth choices in our implementation of local linear regression here.}---which uses the outcome variable $Y$ and running variable $S$ to produce ``robust nonparametric confidence intervals'' for the ATE at the cutoff. Meanwhile, the second row of Table \ref{tab:llrEstimates} uses the approach of \cite{calonico2019regression}---also implemented in the \texttt{rdroubst} \texttt{R} package---to adjust for the additional covariates $\mathbf{X}$ in the RDD. In either case, local linear regression methods estimate that the treatment effect is negative---i.e., that the grant lowers dropout rate---but this effect is not statistically significant. This finding holds for units \textit{at the cutoff} of 15,000 euros, but it does not assess whether or not there is a treatment effect beyond the cutoff in this RDD. Now we will turn to local randomization approaches to assess this.

\begin{table}
	\centering
	\begin{tabular}{|c|c|c|}
	\hline
		\textbf{Method} & \textbf{Point Estimate} & \textbf{Confidence Interval} \\
	\hline
	LLR & -0.153 & (-0.321, 0.015) \\
	LLR (Covariates) & -0.110 & (-0.259, 0.017) \\
	\hline
	\end{tabular}
	\caption{Estimate of the ATE at the cutoff using local linear regression (LLR) and LLR using additional covariates.}
	\label{tab:llrEstimates}
\end{table}

\subsection{Analyzing the Data using Local Randomization}

The local linear regression approach used in Section \ref{ss:llrAnalysis} viewed the running variable $S$ as fixed and provided an estimate for the ATE specifically at the cutoff of 15,000 euros. In contrast, the local randomization approach posits that a hypothetical randomized experiment around the cutoff has occurred due to stochasticity in $S$. Assuming that $S$ is stochastic is particularly realistic for this application, because---as discussed in \citet[Section 5]{li2015evaluating}---it is a complicated measure of income computed by fiscal experts, and students do not have any ability to manipulate their measurement of $S$.

In what follows, we will consider different hypothetical randomized experiments that may have occurred within a window $\mathcal{U}_h$ around the cutoff, due to this stochasticity in $S$. Each hypothetical experiment is characterized by a bandwidth $h$ and an assignment mechanism $P(\mathbf{Z} | \mathbf{X})$. After these are specified, we will estimate the ATE for units in $\mathcal{U}_h$; this is in contrast with local linear regression approaches, which estimate the ATE for units at the cutoff.

First, in Section \ref{sss:completeRandAnalysis}, we will focus on Complete Randomization---defined in (\ref{eqn:completeRandomizationNT})---which is the most common assignment mechanism in the local randomization literature \citep{sales2014limitless,cattaneo2015randomization,matteiMealli2016,cattaneo2017comparing}. Then, in Section \ref{sss:blockRandAnalysis}, we will focus on Block Randomization, defined in (\ref{eqn:blockRandomization}). To our knowledge, Block Randomization has not been used in an application of local randomization for RDDs, but (as we have discussed throughout) its use is a straightforward extension of previous local randomization methods. In both sections, we will show how to (1) use methods from Section \ref{ss:windowProcedure} for selecting the window $\mathcal{U}_h$, (2) use methods from Section \ref{s:estimation} conduct inference for the local ATE within this window, and (3) estimate the ATE across several ``plausible'' bandwidths to assess if inferences are sensitive to the bandwidth choice.

\subsubsection{Analyzing the Data Assuming Complete Randomization} \label{sss:completeRandAnalysis}

\begin{figure}
	\centering
	\includegraphics[scale=0.7]{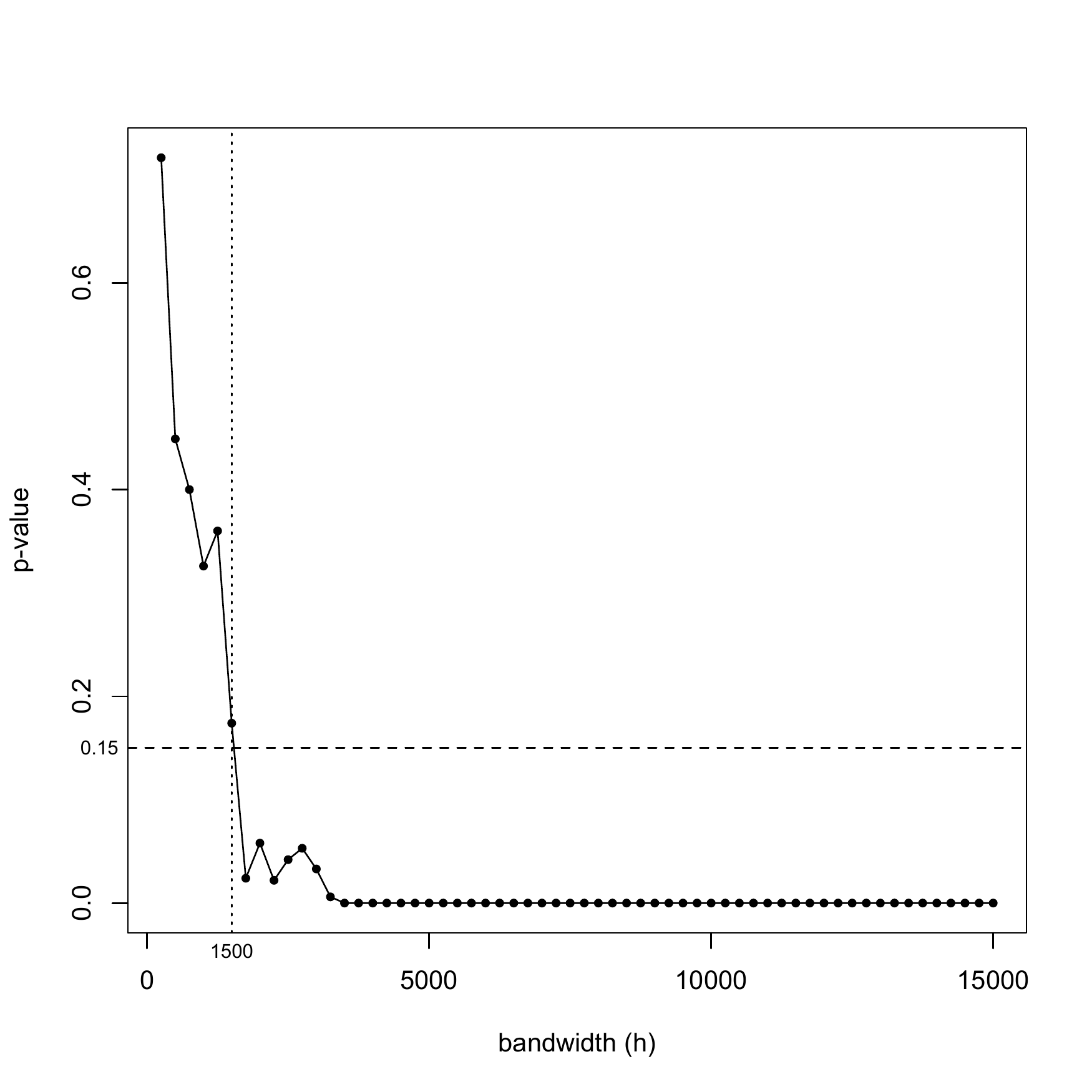}
	\caption{Randomization-based $p$-values for covariate balance using the Mahalanobis distance across bandwidths $h$.}
	\label{fig:mdPlot}
\end{figure}

For this section, assume that we would like to find a window $\mathcal{U}_h = \{i: s_0 - h \leq S \leq s_0 + h\}$ such that Complete Randomization plausibly holds. The running variable ranged from 1 to 20,000 euros in the dataset. The procedure from Section \ref{ss:windowProcedure} outlines how to select $\mathcal{U}_h$. In particular, we will do the following:
\begin{enumerate}
	\item Choose a bandwidth $h$.
	\item Permute $Z^{obs}$ many times for units in $\mathcal{U}_h$.
	\item For each permutation, compute the Mahalanobis distance\footnote{The Mahalanobis distance\citep{mahalanobis1936generalised}, defined as $\frac{N_T N_C}{N} \left( \overline{X}_T - \overline{X}_C \right)^T \left[ \text{cov}(\mathbf{X}) \right]^{-1} \left( \overline{X}_T - \overline{X}_C \right)$, acts as a multivariate analog to the standardized covariate mean difference.} between the treatment and control covariates.
	\item Compute the randomization-based $p$-value---defined in (\ref{eqn:randPValue}).
\end{enumerate}
We used the Mahalanobis distance because it acts as a global measure of balance and is commonly used for assessing covariate balance in observational studies \citep{gu1993comparison,stuart2010matching,diamond2013genetic,branson2019evaluating}. Figure \ref{fig:mdPlot} shows the randomization-based $p$-value from the above procedure across different bandwidths $h \in \{250, 500,\dots, 14750, 15000\}$. Using the rule-of-thumb from \cite{cattaneo2015randomization}, we can assume Local Complete Randomization for the largest bandwidth such that the $p$-value is at least 0.15. From Figure \ref{fig:mdPlot}, we can see that the $p$-value is greater than 0.15 for $h \leq 1500$. This is consistent with \cite{li2015evaluating}, who also concluded---using a Bayesian testing procedure---that covariates are adequately balanced for $h \leq 1500$.

\begin{table}
	\centering
	\begin{tabular}{|c|c|c|}
	\hline
		\textbf{Bandwidth} & \textbf{Point Estimate} & \textbf{Confidence Interval} \\
	\hline
	\hline
	$h = 1500$ & -0.051 & (-0.136, 0.034) \\
	$h = 1000$ & -0.121 & (-0.229, -0.012)  \\
	$h = 500$ & -0.066 & (-0.217, 0.085) \\
	\hline
	\end{tabular}
	\caption{Estimates and confidence intervals for the ATE under Complete Randomization for several bandwidths.}
	\label{tab:crEsts}
\end{table}

Thus, it seems reasonable to estimate the ATE under Complete Randomization for any $h \leq 1500$. Table \ref{tab:crEsts} shows point estimates and confidence intervals for the ATE under Complete Randomization for $h \in \{500, 1000, 1500\}$ using results from Section \ref{ss:howToCR}. Interestingly, this analysis finds that the grant significantly reduces the average dropout rate for students within 1,000 euros of the cutoff but not for those within 500 euros. To better assess this peculiar finding, we will consider variations of Local Block Randomization for this dataset.

\subsubsection{Analyzing the Data Assuming Block Randomization} \label{sss:blockRandAnalysis}

Now for this section, assume that we would like to find a window $\mathcal{U}_h = \{i: s_0 - h \leq S \leq s_0 + h\}$ such that Block Randomization plausibly holds. All of the covariates (other than high school grade) are categorical, and there are $2 \cdot 5 \cdot 3 \cdot 2 \cdot 6 = 360$ different combinations for these covariates. High school grades are integers that range from 60 to 100, and thus there are $360 \cdot 41 = 14,760$ possible combinations for all covariates.

The most precise type of hypothetical block randomized experiment defines the blocks in (\ref{eqn:blockRandomization}) as these 14,760 possible combinations. Unsurprisingly, we were not able to find large enough subsets of treatment and control units that exactly match on all covariates to viably conduct causal inferences. A similar difficulty occurred when we attempted to match on all covariates except high school grade. This is in line with the well-known ``curse of dimensionality'' of exact matching; i.e., it is typically not plausible to exactly match on many covariates in an observational study \citep{stuart2010matching,iacus2012causal,visconti2017handling}.

Instead, we matched on a few covariates (year, sex, and university) which each only have two or three categories. We will consider several variants of Local Block Randomization, where we block on one or more of these three covariates. The ``Block Type'' column of Table \ref{tab:brEsts} shows the seven different variants of Local Block Randomization that can be considered for these covariates. 

Before we conduct an analysis assuming a particular variant of Local Block Randomization, first we must assess if that variant holds for a particular window $\mathcal{U}_h$ around the cutoff. In the previous section, we found that Local Complete Randomization was plausible for $h \leq 1500$, so we will assess if Local Block Randomization holds for the bandwidths $h \in \{500, 1000, 1500\}$. To do this, we again use the randomization test procedure discussed in Section \ref{ss:windowProcedure}, but assuming these variants of Local Block Randomization instead of Local Complete Randomization. The only thing that changes when implementing the randomization test is that, instead of simply permuting $Z^{obs}$ across all units, we only permute $Z^{obs}$ within the blocks defined by the particular variant of Local Block Randomization. Then, as before, we compute the Mahalanobis distance across these permutations, and we compare the resulting distribution of Mahalanobis distances to the observed Mahalanobis distance to compute the randomization test $p$-value.

The ``Rand. Test $p$-value'' column of Table \ref{tab:brEsts} shows the resulting $p$-values for each variant of Local Block Randomization. Generally speaking, Block Randomization appears to hold for different windows (in the sense that the randomization test $p$-values tend to be greater than 0.15) except when we only block on year and/or sex for $h = 1500$. The Local Block Randomization point estimates and confidence intervals for $h \in \{500, 1000, 1500\}$ are shown in Table \ref{tab:brEsts}.

\begin{table}
	\centering
	\begin{tabular}{|c|c|c|c|}
	\hline
		\textbf{Block Type} & \textbf{Rand. Test $p$-value} & \textbf{Point Estimate} & \textbf{Confidence Interval} \\
	\hline
	\hline
	\textit{Year} & & & \\
	\sout{$h = 1500$} & 0.126 & \sout{-0.061}  & \sout{(-0.120, -0.002)}  \\
	$h = 1000$ & 0.311 & -0.098  & (-0.175, -0.022)   \\
	$h = 500$ & 0.324 & -0.089  & (-0.194, 0.017)  \\
	\hline
		\textit{Sex} & & & \\
	\sout{$h = 1500$} & 0.138  & \sout{-0.054}   & \sout{(-0.139, 0.032)}   \\
	$h = 1000$ & 0.255  & -0.131   & (-0.241, -0.021)    \\
	$h = 500$ & 0.411  & -0.075   & (-0.229, 0.078)  \\
	\hline
			\textit{Uni.} & & & \\
	$h = 1500$ & 0.592  & -0.026   & (-0.117, 0.065)   \\
	$h = 1000$ & 0.276  & -0.111   & (-0.225, 0.002)    \\
	$h = 500$ & 0.435  & -0.065   & (-0.228, 0.097)  \\
	\hline
		\textit{Year \& Sex} & & & \\
	\sout{$h = 1500$} & 0.064 & \sout{-0.061}  & \sout{(-0.121, -0.002)}  \\
	$h = 1000$ & 0.275 & -0.103  & (-0.180, -0.025)   \\
	$h = 500$ & 0.245 & -0.097  & (-0.206, 0.013)  \\
	\hline
			\textit{Year \& Uni.} & & & \\
	$h = 1500$ & 0.526 & -0.059  & (-0.119, 0.001)  \\
	$h = 1000$ & 0.241 & -0.097  & (-0.175, -0.020)   \\
	$h = 500$ & 0.297 & -0.085  & (-0.193, 0.022)  \\
	\hline
			\textit{Sex \& Uni.} & & & \\
	$h = 1500$ & 0.506  & -0.025   & (-0.117, 0.067)   \\
	$h = 1000$ & 0.250  & -0.117   & (-0.234, -0.000)   \\
	$h = 500$ & 0.424  & -0.070   & (-0.239, 0.099)  \\
	\hline
				\textit{Year \& Sex \& Uni.} & & & \\
	$h = 1500$ & 0.406 & -0.055  & (-0.117, 0.006)  \\
	$h = 1000$ & 0.199 & -0.098  & (-0.177, -0.019)   \\
	$h = 500$ & 0.239 & -0.087  & (-0.199, 0.026)  \\
	\hline
	\end{tabular}
	\caption{Randomization $p$-values, point estimates, and confidence intervals under variations of Block Randomization for several bandwidths. Analyses are crossed out for cases where $p$-value $\leq 0.15$.}
	\label{tab:brEsts}
\end{table}

The confidence intervals under Local Block Randomization are notably more precise than the confidence intervals under Local Complete Randomization, reflecting the gains in this more precise hypothetical experimental design. However, the analyses across different Local Block Randomization assumptions are quite similar to one another. In particular, the analyses corresponding to blocking on (1) year and sex, and (2) year and university, are quite similar to the analysis corresponding to blocking on year alone. This is likely because the year covariate turned out to be moderately related to the outcome ($R^2 = 0.582$ for the linear regression of the outcome on year when $h = 1500$), while the sex and university covariates are only weakly related to the outcome ($R^2 = 0.001$ and $R^2 = 0.020$ for the linear regression of the outcome on sex and on university, respectively). Thus, assuming Local Block Randomization (or some other locality assumption) only results in precision gains when one can condition on covariates that are related to the outcome, which is unsurprising given the long understanding that block randomization is most beneficial if one blocks on relevant covariates (e.g., see the experimental design textbooks \cite{cox2000theory}, \cite{box2005statistics}, and \cite{seltman2012experimental}). However, as discussed in Section \ref{s:window}, it is important that no outcome information is used to choose locality assumptions on the assignment mechanism, in order to avoid biasing results \citep{rubin2008objective}.

Within the context of the application, there appears to be a negative treatment effect, indicating that the university grant tends to decrease dropout rates. The confidence intervals for all analyses (though narrower than the Complete Randomization confidence intervals) tend to widen as $h$ decreases, due to the reduction in sample size when $h$ decreases. These findings are fairly consistent across different variants of Local Block Randomization. However, for $h = 1500$, we may have some doubts in these findings, due to the lack of Block Randomization for the year and sex covariates for this bandwidth. This demonstrates how positing different assignment mechanisms within a single RDD can provide more nuanced sensitivity analyses as well as more precise inferences for causal effects. The findings in Table \ref{tab:brEsts} also show how, similar to \cite{li2015evaluating} and \cite{calonico2019regression}, we can adjust for additional covariates in an RDD, but (unlike \cite{li2015evaluating} and \cite{calonico2019regression}) without placing any modeling assumptions on the outcome, instead leveraging locality assumptions on the assignment mechanism.

We could extend our methodology to include model-based adjustments on covariates that we did not block on through locality assumptions, and this could be done by specifying models for the outcome as well as the compliance behavior (e.g., using methods from \citet[Chapter 25]{imbens2015causal}). We could also include model-based adjustments on the running variable. For example, the validity of our analysis depends on Local Strong Ignorability---defined in (\ref{eqn:localStrongIgnorability})---holding for the $\mathcal{U}_h$ chosen in our analysis, and \cite{matteiMealli2016} discuss how model-based adjustments on the running variable may be helpful if this assumption is suspect. However, by avoiding model specifications, our point estimate retains the transparency of a simple mean-difference estimator---a transparency that is often lost in complex models \citep{lin2013agnostic}.

Finally, it is important to contrast the analyses conducted throughout this section with the local linear regression analyses conducted in Section \ref{ss:llrAnalysis}. The methods presented in Section \ref{ss:llrAnalysis} only estimate the ATE for students \textit{at the cutoff} of 15,000 euros. Meanwhile, the local randomization methods presented in this section estimate the ATE for students whose income is within $h$ of the cutoff. For example, our analysis for $h = 1000$ estimates the ATE for students whose income is in $(14000, 16000)$ euros. Thus, local randomization methods provide a way to estimate causal effects for a more general population, which is particularly useful for policy-relevant contexts like the Italian grant application.

\section{Discussion and Conclusion} \label{s:conclusion}

Regression discontinuity designs (RDDs) are a common quasi-experiment in economics, education, medicine, and statistics. The most popular methodologies for estimating causal effects in an RDD rely on continuity assumptions, view the running variable as fixed, and focus estimation on the ATE at the cutoff. In contrast, a recent strand of literature has developed a local randomization framework for RDDs, which views the running variable as stochastic, thereby introducing randomness to the assignment mechanism. Two benefits of the local randomization framework are that (1) it focuses estimation on the ATE for a subset of units around the cutoff, instead of just for units at the cutoff; and (2) it does not require any modeling assumptions, and instead only requires assumptions about the assignment mechanism.

In this paper, we provided a review of the local randomization literature for RDDs, thereby showing practitioners how they can use methods besides local linear regression to estimate causal effects for a general population within an RDD. The local randomization literature has focused primarily on the Local Complete Randomization assignment mechanism, which is characterized by random permutations of the treatment indicator. Assuming Local Complete Randomization is a strong assumption, because it posits that the propensity scores are equal for all units near the cutoff. In our review, we extended the local randomization framework to allow for any assignment mechanism, such as Bernoulli trials and block randomization, where propensity scores are allowed to differ. This involved outlining an approach for finding a window around the cutoff such that a particular assignment mechanism holds, as well as developing a methodology for estimating causal effects after a window and assignment mechanism have been chosen. For example, we outlined how to analyze a fuzzy RDD assuming Local Block Randomization. This is a new contribution, but it is also a straightforward extension of previous local randomization approaches. Thus, we believe our review shows practitioners how they can further extend the local randomization approach to complex assignment mechanisms that may be suitable for their application.

In this vein, we applied our methodology to a fuzzy RDD that assesses the effects of financial aid on college dropout rates in Italy. This RDD was originally analyzed by \cite{li2015evaluating} assuming Local Complete Randomization and using a Bayesian model that adjusted for covariates and addressed units' compliance behavior in the fuzzy RDD. In contrast, we analyzed this RDD assuming Local Block Randomization, which allows units' propensity scores to differ across blocks defined by covariates. We concluded that financial aid led to a significant decrease in college dropout rates, which is in line with the original findings of \cite{li2015evaluating}. More generally, our application demonstrated how positing different assignment mechanisms within a single RDD can provide more nuanced sensitivity analyses as well as more precise inferences for causal effects. Furthermore, this application showed how our local randomization framework provides a way to make covariate adjustments in RDDs without modeling assumptions.

As we discussed in Section \ref{s:realDataAnalysis}, our methodology could be extended to allow for modeling adjustments in conjunction with locality assumptions on the assignment mechanism. For example, \cite{li2015evaluating}, \cite{cattaneo2017comparing}, and \cite{branson2019nonparametric} show how more complex models can be combined with the Local Complete Randomization assumption. Our review suggests a promising line for future research that explores how different modeling choices from the local linear regression literature and locality assumptions from the local randomization literature can be combined to produce precise causal inferences for RDDs.  

\newpage

\bibliography{localRDDBib}

\bibliographystyle{apa-good}

\end{document}